\documentclass[%
 reprint,
 amsmath,amssymb,
 aps,
]{revtex4-2}

\usepackage{graphicx}
\usepackage{dcolumn}
\usepackage{bm}

\begin{document}

\preprint{APS/123-QED}

\title{Polarization-based cyclic weak value metrology for angular-velocity measurement}
\thanks{A footnote to the article title}%

\author{Zi-Rui Zhong}
\author{Yue Chen}
\author{Wei-Jun Tan}
\author{Xiang-Ming Hu}
\author{Qing-Lin Wu}%
 \email{qlwu@ccnu.edu.cn}
\affiliation{%
\emph{Department of Physics, Central China Normal University, Wuhan 430079, China}
}%
\affiliation{%
\emph{Key Laboratory of Quark \& Lepton Physics (MOE) and Institute of Particle Physics, Central China Normal University, Wuhan 430079, China}
}%

\date{\today}

\begin{abstract}

Weak measurement has been proven to amplify the detection of changes in meters while discarding most photons due to the low probability of post-selection. Previous power-recycling schemes enable the failed post-selection photons to be repeatedly selected, thus overcoming the inefficient post-selection and increasing the precision of detection.
In this study, we focus on the polarization-based weak value angular-velocity measurement and  introduce three cyclic methods to enhance the accuracy of detecting time shift in a Gaussian beam: power recycling, signal recycling, and dual recycling schemes. By incorporating one or two partially transmitting mirrors into the system, both the power and signal-to-noise ratio (SNR) of the detected light are substantially enhanced.
Compared to non-polarization schemes, polarization-based approaches offer several advantages, including lower optical loss, unique cyclic directions, and a wider optimal region. These features effectively reduce crosstalk among different light paths and theoretically eliminate the walk-off effect, thus yielding improvements in both theoretical performance and application.

\end{abstract}

\maketitle


\section{Introduction}
Since first introduced by Aharonov, Albert and Vaidman in Ref.\cite{1}, weak measurement has shown its numerous potentials in various precise measurements. Unlike the classical (or strong) measurements set forth by von Neumann\cite{2}, weak measurement involves a significantly weak coupling between the probe and the system, permitting a small change of coupling parameter to be converted into a large change in a meter variable\cite{1,3}. Consequently, it can be used to reconsider some interesting quantum phenomena such as Hardy’s paradox\cite{4,5,6,7}, three-box problem\cite{8,9,10} and quantum Cheshire Cats\cite{11,12,13,14,15}.
By appropriately preparing pre- and post-selected states, weak measurement enables the determination of a "weak value" that encapsulates information regarding the weak interaction process. Generally denoted as $A_w=\left\langle f\right|\hat{A}  \left | i  \right \rangle /\left \langle f  | i  \right \rangle $, where $\left | i  \right \rangle $, $\left | f  \right \rangle$ are the pre- and post-selected states, respectively, and $\hat{A}$ represents the measured observable.
Notably, due to the presence of $\left \langle f  | i  \right \rangle$ in the denominator, $A_w$ can be very large if $\left | i  \right \rangle $ and $\left | f  \right \rangle$ are nearly orthogonal. Thus, it has the potential to detect many small physical effects such as the spin Hall effect\cite{16,17,18}, Goos-Hänchen shift\cite{19,20}, beam deflection\cite{21}, velocity\cite{22}, phase shift\cite{24,25}, temperature\cite{26}, angular-velocity\cite{27,28,29} and resonance\cite{30}, to name a few.

Theoretical analysis has pointed out that weak measurement can outperform the conventional measurement in the presence of detector saturation and pixel noise\cite{PhysRevLett.118.070802}. 
Besides, it has been proven that weak measurement can suppress technique noise in some circumstances\cite{PhysRevA.80.041803,16}, and can even yield several orders of magnitude improvement over conventional measurements through imaginary weak-value measurements\cite{PhysRevLett.105.010405,PhysRevLett.107.133603,PhysRevA.85.060102,PhysRevX.4.011031}. Furthermore, some reports even proposed Heisenberg-Scaling precision post-selection measurement using coherent states and photon-counting detection\cite{PhysRevLett.114.210801,PhysRevLett.121.060506,PhysRevA.106.022619}, which challenges the necessity of entanglement in quantum-enhanced precision. Conversely, negative discussions have primarily centered around the significant loss of photons due to low successful post-selection probabilities, resulting in a considerable reduction in the attainable Fisher information\cite{PhysRevX.4.011031,PhysRevLett.112.040406,PhysRevLett.114.210801,PhysRevX.4.011032}. This delicate balance has sparked controversial debates in previous literature\cite{PhysRevLett.112.040406,PhysRevLett.114.210801,PhysRevX.4.011032}. 
To address this issue, recycling techniques have been proposed as they are highly compatible with weak measurement and offer the potential to optimize the prevalent disadvantage of diminished Fisher information resulting from low post-selection probabilities. At present, three types of weak-value-based recycling techniques have been proposed: power-recycling, signal-recycling and dual-recycling.
The power-recycling technique\cite{31,32,33,34,35}, proposed by introducing a partially transmitting mirror (PTM) at the bright port of an interferometer, offers a approach for reusing failed post-selection photons. In ideal conditions, this technique enables the detection of all input light, thus maximizing the efficiency of the system. The similar conclusion is obtained for the signal-recycling weak measurement\cite{36}, which works by placing the PTM at the dark port of the interferometer. Furthermore, these power-recycling and signal-recycling techniques can be combined within a dual-recycling scheme to achieve an improved optimal region\cite{37,38,39,zhong2023dualrecycled}. 

Previous dual-recycled interferometric weak-value-amplification (WWA) setup obtains large precision improvement while sacrificing some of WWA effect of pointer due to the walk-off effect. In addition, the intricate path of cyclic photons within the interferometer gives rise to inevitable crosstalk, thereby increasing system loss. 
To address these challenges, we propose a cyclic scheme for polarization-based weak-value amplification, building upon the angular-velocity measurement framework presented in Ref. \cite{40} . In contrast to the non-polarization cyclic schemes, we substitute the polarization-beam-splitter (PBS) for beam-splitter(BS). This modification simplifies the light path to be exclusively clockwise $\circlearrowright$ and reduces the optical loss.  Moreover, the unidirectional cyclic paths permit a filter to refresh all cyclic photons prior to their final weak interaction, eliminating the walk-off effect.

\begin{figure*}[t]
\centering
\includegraphics[trim= 0 0 0 0 ,clip, scale=0.58]{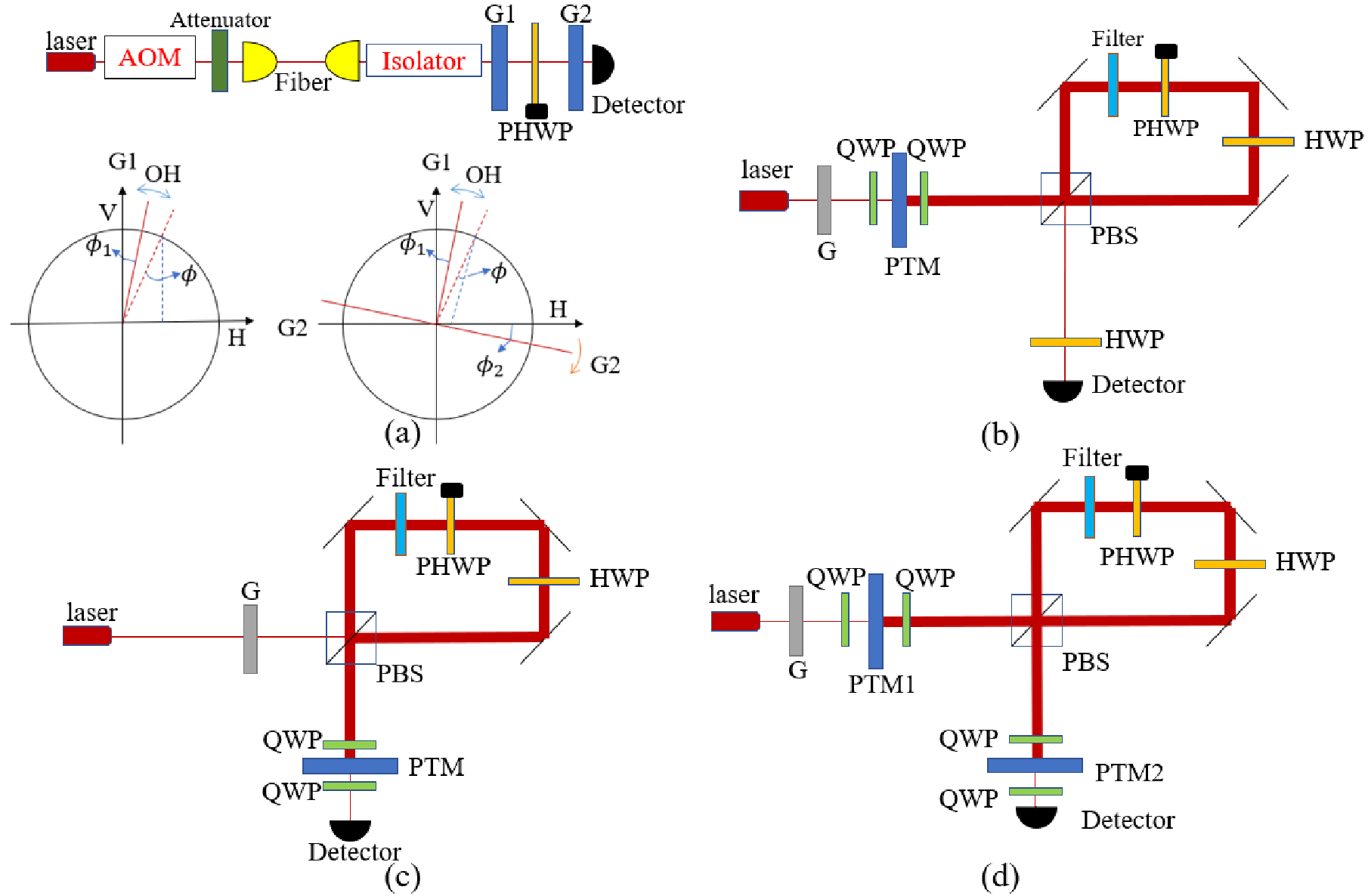}
\caption{(Color online) (a)Schematic of weak-value-based angular-velocity measurement. A laser wave is generated by an acoustic optical modulator (AOM) and enters a polarization-dependent angular velocity measurement system consisting of two Glan prisms (G1 and G2) and the PHWP. The small angular velocity $\omega$ is induced by the PHWP and finally measured by the detector. The two subfigures below show the different orientations of the optical axis of the Glan prisms and the HWP (b): The power-recycling scheme. The PBS, which distinguishes the polarization between H and V, is combined with the PTM to reuse the failed post-selection photons repeatedly. Two QWPs near the PTM provide the initial polarization and rotate the polarization direction of cyclic light back to V. The filter in front of the PHWP refreshes the beam profile on each pass. The HWP rotates the polarization of light by $\pi/2$. (c)The signal-recycling scheme. The PTM at the output port is combined with the WWA system to form a signal-recycling cavity, thus improving the detected signal. (d)The dual-recycling scheme. Combining the power- and signal-recycling PTMs into one scheme further enhances the precision of detection. QWP: quarter-wave plate. HWP: half-wave plate. PBS: polarization-beam-splitter. PHWP: piezo-driven half-wave plate. PTM: partially transmitting mirror. H: horizontal. V: vertical. OH: the optical axis of the HWP}
\label{Fig.1}
\end{figure*}

\section{Standard WWA setup}
We first review the standard WWA setup for angular-velocity measurement in Ref. \cite {40}. As shown in Fig. 1(a), a non-Fourier limit Gaussian pulse $I_0\left(t\right)=\left(N^2/2\pi\tau^2\right)^{1/2}\exp\left(-t^2/2\tau^2\right)$ , where $N$ is the number of photons and $\tau$ is the length of pulse, is sent to a polarization-dependent system. The first Glan prism (G1) combined with the half-wave plate (HWP) provides the pre-selected state and the second Glan prism (G2) provides the post-selection.
In this system, the optical axis of G1 is vertical and the angle between G1 and HWP is $\phi_1$. The Gaussian pulse acts as a probe, and the weak interaction is induced by the piezo-driven half-wave plate (PHWP) with a angular-velocity $\omega$. After the interaction, the joint state becomes

\begin{equation}
    \begin{aligned}
    \left | \Psi  \right \rangle =
    &\left [ \sin \left ( 2\phi_1+2\omega t \right ) \left | H  \right \rangle +\cos \left ( 2\phi_1+2\omega t \right )\left | V  \right \rangle   \right ]\\
    &(\frac{N^2}{2\pi\tau^2} )^{\frac{1}{4} } e^{-\frac{t^2}{4\tau^2}  } \left | t  \right \rangle\\
    =\frac{i}{\sqrt{2} }&  [ e^{-2i(\phi_1+\omega t)}\left | R  \right \rangle  - e^{2i(\phi_1+\omega t)}\left | L  \right \rangle]\left|\varphi_0\right\rangle,\\
    \end{aligned}    
\end{equation}
where we use the circularly polarized states $\left | R  \right \rangle =\frac{1}{\sqrt{2} } \left ( \left | H  \right \rangle -i\left | V  \right \rangle  \right ) $ and $\left | L  \right \rangle =\frac{1}{\sqrt{2} } \left ( \left | H  \right \rangle +i\left | V  \right \rangle  \right ) $ as basis system states, and use $\left|\varphi_0\right\rangle$ to express the initial state of probe $\left|\left.\varphi_0\right\rangle\right.=\sqrt{I_0\left(t\right)}\left|\left.t\right\rangle\right.=\left(N^2/2\pi\tau^2\right)^{1/4}\exp\left(-t^2/4\tau^2\right)\left|\left.t\right\rangle\right.$. Compared with the initial joint state (after the G1) 
$\left | \Psi_0  \right \rangle = \left | V  \right \rangle\left|\varphi_0\right\rangle=\frac{i}{\sqrt{2} }  ( \left | R  \right \rangle  - \left | L  \right \rangle)\left|\varphi_0\right\rangle$, the piezo-driven half-wave-plate (PHWP) here acts as introducing a $(4\omega t+4\phi_1)$ phase shift between the left and right circularly polarized components of light, including a $4\omega$ frequency shift induced by the weak interaction.
Thus, we use ${\hat{U}}_w=\exp\left(i2\omega t\hat{A}\right)$, where $\omega$ is angular velocity by rotating HWP and $\hat{A}$ is a Hermitian operator $\hat{A}=\left|\left.L\right\rangle\right.\left\langle\left.L\right|\right.-\left|\left.R\right\rangle\right.\left\langle\left.R\right|\right.$, to express the weak interaction and ${\hat{U}}_\phi=\exp\left(i2\phi_1\hat{A}\right)$ to represent the polarization rotation $\phi_1$ produced by the PHWP. The pre-selected state of system is $\left|\left.\psi_{pre}\right\rangle\right.={\hat{U}}_\phi\left|\left.V\right\rangle\right.=\frac{i}{\sqrt2}\left[\exp\left(-i2\phi_1\right)\left|\left.R\right\rangle-\exp\left(i2\phi_1\right)\left|\left.L\right\rangle\right.\right.\right]$ and the post-selected state is given by the second Glan prism (G2), 
$\left.\left|\psi_{pos}\right.\right\rangle=\cos{\phi_2\left|\left.H\right\rangle\right.}-\sin{\phi_2\left.\left|V\right.\right\rangle}=\frac{i}{\sqrt2}\left[\exp\left(-i2\phi_2\right)\left|\left.R\right\rangle+\exp\left(i2\phi_2\right)\left|\left.L\right\rangle\right.\right.\right]$, where the initial optical axis of the G2 is horizontal and $\phi_2$ is the angle of rotation. [see Fig.1 (a)]. Therefore, the intensity of detected light is 
\begin{equation}
    \begin{aligned}
    I_d\left(t\right)
    & =\left|\left\langle\psi_{pos}\right|\hat{U}_w \left|\psi_{pre} \right \rangle \left|\left.\varphi_0\right\rangle\right.\right|^2\\
    & \approx\frac{N}{\sqrt{2\pi\tau^2}}{\sin}^2\phi \exp\left[-\frac{1}{2\tau^2}\left(t-\frac{4\omega\tau^2}{\phi}\right)^2\right],
    \end{aligned}
\end{equation}
where we introduce the angle $\phi=2\phi_1-\phi_2$ and assume $2\omega\tau\ll\phi\ll1$.  The corresponding weak value is given by $A_w=\left\langle \psi_{pos}\right|\hat{A}  \left | \psi_{pre}  \right \rangle /\left \langle \psi_{pos}  | \psi_{pre} \right \rangle \approx -i/\phi$. Compared with the incident light, the time shift induced by the PHWP is $\delta t=\frac{4\omega\tau^2}{\phi}=4\omega\tau^2\left|A_w\right|$, which is related to the weak value. Based on Fisher information (FI) theory\cite{42,43}, the FI of $\delta t$ is obtained from the detected light $I_d(t)$, 
\begin{equation}
    \mathcal{F}\left(\delta t\right)=\int{dtI_d(t)\left|\frac{d}{d\delta t}lnI_d(t)\right|^2}\approx\frac{N\phi^2}{\tau^2}.
    \label{e2}
\end{equation}
So the minimum uncertainty of time shift determined by Cramér-Rao bound (CRB) satisfies
\begin{equation}
    \Delta (\delta t)=\frac{1}{\sqrt{F(\delta t)}}=\frac{\tau}{\phi\sqrt{N}}.
    \label{e3}
\end{equation}
Thus, the minimum uncertainty of angular velocity is
\begin{equation}
    \Delta \omega=\frac{\phi}{4\tau^2} \Delta\left ( \delta t \right ) \approx\frac{1}{4\sqrt[]{N}\tau  }.
    \label{e4}
\end{equation}
Therefore, the corresponding SNR is given by
\begin{equation}
    \rm{SNR}=\frac{\omega}{\left(\Delta \omega\right)}=4\omega\tau\sqrt N=\frac{\sqrt N\phi}{\tau}\delta t.
    \label{e5}
\end{equation}

\begin{figure*}[t]
\centering
\includegraphics[trim= 0 0 0 0 ,clip, scale=0.68]{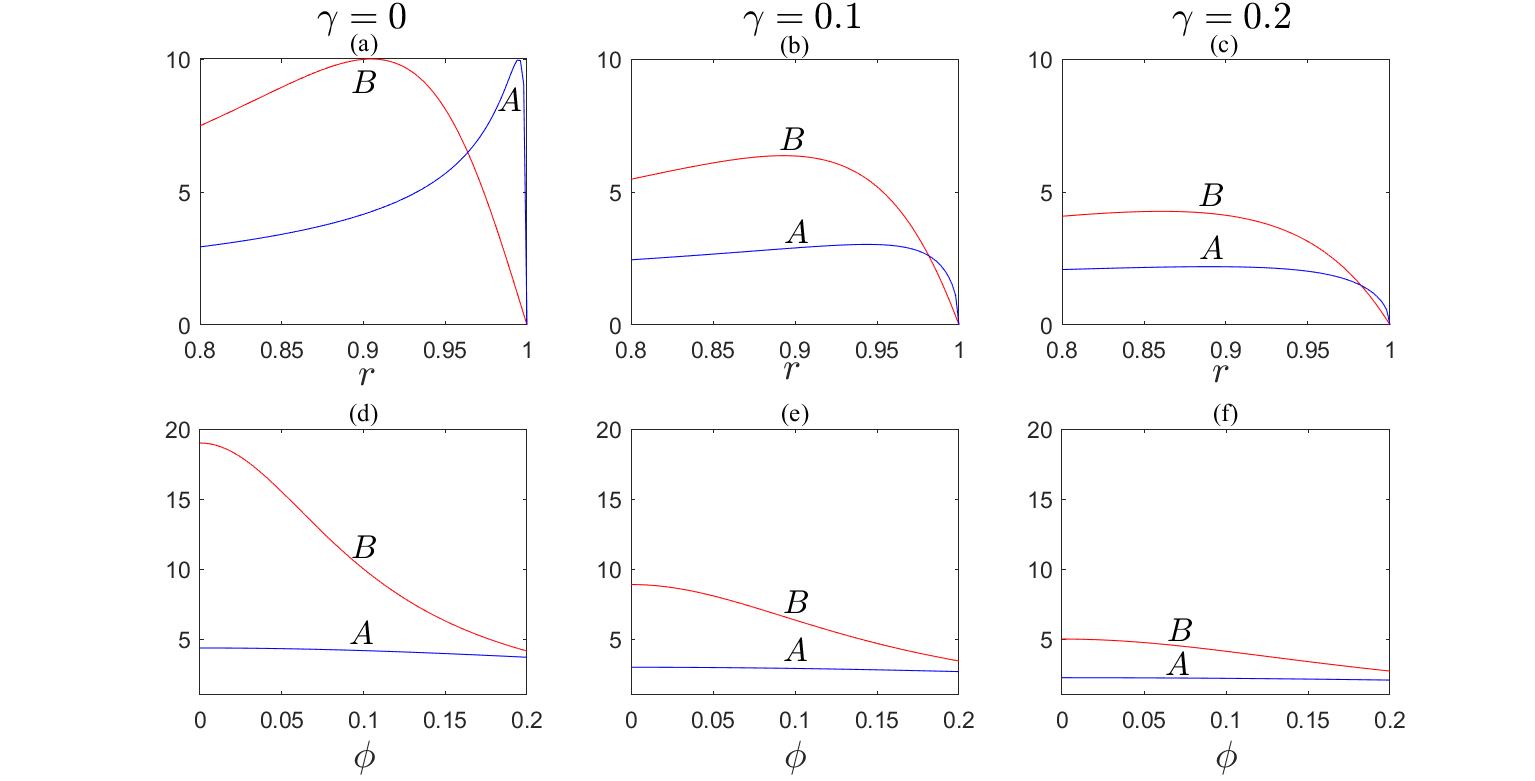}
\caption{(Color online) Comparison of power-, signal- and dual-recycling schemes. (a), (b) and (c) correspond to the case of A and B varying with $r$ under different values of $\gamma$: 0, 0.1 and 0.2 where $\phi=0.1$. In (d), (e) and (f), we assume $r=0.9$ and plot A, B varying with $\phi$ under $\gamma=0,\ 0.1,\ 0.2$, respectively. A and B: the improvement factor of power (or signal) and dual recycling schemes. $r$: the reflection coefficient of PTM. $\gamma$: optical loss.}
\label{Fig.2}
\end{figure*}

\section{Recycling technique}
The methodologies of recycling techniques refer to references \cite{31} and \cite{zhong2023dualrecycled}, where the improvement of precision is originated from the increasing of weak-value related photons being detected. The main difference in this method is that it effectively eliminates the opposite traverse shift of the recycling profile by utilizing a filter in a manner made possible by the use of a polarization-beam-splitter (PBS). This PBS adjusts the optical path within the recycling loop, thereby mitigating the walk-off effect. In additions, the Gaussian light here is modulated in the time-domain as opposed to the traditional $x$-domain. Such arrangement induces a $\omega$-related time shift, which enables higher-precision detection. These changes distinguish this model from previous works.

\subsection{Power-recycling}
The power-recycled weak-value setup is shown in Fig. 1(b). The initial state $\left|\left.V\right\rangle\right.$ is provided by the Glan prism (G) and two QWPs, and we use the combination ‘HWP1 PBS HWP2’ to replace the G2 for the post-selection. The PTM, whose reflection and transmission coefficients are $r$ and $p$ $\left ( r^2+p^2=1 \right )$, is placed between two QWP, thus reflecting the failed post-selected light while rotating the light polarization to $\left|\left.V\right\rangle\right.$. Here we define two orthogonal states $\left|\left.\psi_1\right\rangle\right.$ and $\left|\left.\psi_2\right\rangle\right.$ to represent the input and output system states, where $\left|\left.\psi_1\right\rangle\ =\left|\left.V\right\rangle\right.\right.=\frac{i}{\sqrt2}\left(\left|\left.R\right\rangle-\left|\left.L\right\rangle\right.\right.\right)$ and $\left|\left.\psi_2\right\rangle\ =\left|\left.H\right\rangle\right.\right.=\frac{1}{\sqrt2}\left(\left|\left.R\right\rangle+\left|\left.L\right\rangle\right.\right.\right)$. Post-selected by the input and output ends, the meter states become $\left|\left.\varphi_{ref}\right\rangle\right.=\left \langle V \right |\hat{U_w}\hat{U_\phi }  \left | V  \right \rangle \left|\left.\varphi_0\right\rangle\right.$ and $\left|\left.\varphi_{out}\right\rangle\right.=\left \langle H \right |\hat{U_w}\hat{U_\phi }  \left | V  \right \rangle \left|\left.\varphi_0\right\rangle\right.$ , respectively. This produces two measurement operators $M_{11}=\left \langle \psi_1 \right | {\hat{U}}_w{\hat{U}_\phi}\left | \psi_1  \right \rangle =\cos{\left(\phi+2\omega t\right)}$ and $M_{12}=\left \langle \psi_2 \right | {\hat{U}}_w{\hat{U}_\phi}\left | \psi_1  \right \rangle =\sin{\left(\phi+2\omega t\right)}$. We introduce the non-unitary operator $\hat{L} =\sqrt{1-\gamma}$, where $\gamma$ is the single-pass power loss, to express the loss of optical imperfection in one return. Assuming the length of one traversal is $l_{cav}$, the pulse transition time of per traversal is given by
$t_{cav}=2l_{cav}/c$. Generally, both the measurement operators and the meter state are related to the number of traversals $n$. For example, $M_{11}$ should be written as ${M_{11}}^n=\cos{\left[\phi+2\omega\left(t-nt_{cav}\right)\right]}$. \cite{35}  proved that this change is small and only induces a constant delay, which can be eliminated. Therefore, with the resonance cavity, the amplitude of the detected signal is given by the sum of amplitude from all traversal numbers,
\begin{equation}
    \left|\left.\varphi_p\right\rangle\right.=pM_{12}\sum_{n=0}^{\infty}{\left(rLM_{11}\right)^n\left|\left.\varphi_0\right\rangle\right.}
\end{equation}
It is a summation of the convergence series so that there is a maximum value of $n$, denoted by $n_{max}$. Therefore, the formula above can be simplified as 
\begin{equation}
    \begin{aligned}
    &\left|\left.\varphi_p\right\rangle\right.
    =pM_{12}\sum_{n=0}^{n_{max}}{\left(rLM_{11}\right)^n\left|\left.\varphi_0\right\rangle\right.}\\
    &\approx\left(\frac{N^2}{2\pi\tau^2}\right)^\frac{1}{4}\exp\left(-\frac{t^2}{4\tau^2}\right)\frac{p\sin{\left(\phi+2\omega t\right)}}{1-rL\cos{\left(\phi+2\omega t\right)}}|t\rangle
    \end{aligned}
\end{equation}
Next, we do a Taylor expansion on the function $f\left(t\right)=p\sin{\left(\phi+2\omega t\right)}/\left[1-rL\cos{\left(\phi+2\omega t\right)}\right]$ and make an approximation $f\left(t\right)\approx f\left(0\right)+tf^\prime\left(0\right)\approx \exp\left[-f^\prime\left(0\right)t/f\left(0\right)\right]$. Then the amplitude of detected state is
\begin{equation}
    \left\langle t\right.\left|\left.\varphi_p\right\rangle\right.\approx A\left(\frac{N^2}{2\pi\tau^2}\right)^\frac{1}{4}\sin^2{\left(\phi+2\omega t\right)}\exp\left[-\frac{\left(t-\delta t_p\right)^2}{4\tau^2}\right]
\end{equation}
where
\begin{equation}
    A=\frac{p}{1-r\sqrt{1-\gamma}\cos{2\phi}}
\end{equation}
and
\begin{equation}
    \delta t_p=\frac{2\omega\tau^2\left(\cos{\phi}-r\sqrt{1-\gamma}\right)}{\sin{\phi}\left(1-r\sqrt{1-\gamma}\cos{2\phi}\right)}.
\end{equation}
Due to the walk-off effect, the time shift changes from $\delta t$ to $\delta t_p$. But if placing a filter in front of the PHWP, each time being reflected by the PTM, the light pass through the filter and is projected into $\left|\left.\varphi_0\right\rangle\right.$. This leaves the pre-filter state as $\left|\left.\varphi\prime\right\rangle\right._{pow}=M_{11}\left|\left.\varphi_0\right\rangle\right./\sqrt{\left|M_{11}\left|\left.\varphi_0\right\rangle\right.\right|^2}$. Thus, the probability of surviving the filter is
\begin{equation}
    \begin{aligned}
        p_f=&\left|\left\langle\varphi_0\middle|\varphi\prime\right\rangle\right|^2=\frac{\cos^2{\phi}}{\sinh{4\omega^2\tau^2}+\cos^2{\phi}e^{-4\omega^2\tau^2}}\\
        &\approx1-\phi^2\left(4\omega^2\tau^2\right)-\left(4\omega^2\tau^2\right)^2/2+\cdots,
    \end{aligned}
\end{equation}
where we make the approximation in the weak value range, $2\omega\tau\ll\phi\ll1$. In this way, the time shift is refreshed every cycle, eliminating the walk-off effect while adding a minimum ‘filter’ loss $\gamma_{min}\approx4\omega^2\tau^2\phi^2$ to the system\cite{31}. Therefore, the power of the detected signal is given by 
\begin{equation}
    I_{pow}\approx\left(\frac{N^2}{2\pi\tau^2}\right)^\frac{1}{2}A^2\sin^2{\left(\phi+2\omega t\right)}\exp\left[-\frac{\left(t-\delta t\right)^2}{2\tau^2}\right].
\end{equation}
Similar to the calculation in Eqs. (\ref{e2}-\ref{e5}), the corresponding minimum uncertainty determined by Cramér-Rao bound is 
\begin{equation}
    \Delta \omega_p \approx\frac{1}{4A\sqrt[]{N}\tau  }.
    \label{e6}
\end{equation}
So the corresponding SNR is
\begin{equation}
    {\rm SNR}_{pow}\approx A\frac{\sqrt N\phi}{\tau}\delta t,
    \label{e7}
\end{equation}
which is A times to the SNR of standard weak measurement.

\subsection{Signal-recycling}
The similar methods can be used in the signal-recycled scheme. As shown in Fig. 1(c), the optical axis of Glan prism is vertical, providing the input state $\left|\left.V\right\rangle\right.$. The post-selection is provided by the combination ‘HWP PBS QWPs’. The PTM is placed between the two QWPs to reuse the output signal. This post-selection processes provide two measurement operators: $M_{12}=\left \langle \psi_2 \right | {\hat{U}}_w{\hat{U}_\phi}\left | \psi_1  \right \rangle =\sin{\left(\phi+2\omega t\right)}$ and $M_{22}=\left \langle \psi_2 \right | {\hat{U}}_w{\hat{U}_\phi}\left | \psi_2  \right \rangle =\cos{\left(\phi+2\omega t\right)}$. In this signal-recycled cavity, the amplitude of the detected signal is given by
\begin{equation}
    \begin{aligned}
        &\left|\left.\varphi_s\right\rangle\right.=pM_{12}\sum_{n=0}^{n_{max}}{\left(rLM_{22}\right)^n\left|\left.\varphi_0\right\rangle\right.}\\
        &\approx\left(\frac{N^2}{2\pi\tau^2}\right)^\frac{1}{4}exp\left(-\frac{t^2}{4\tau^2}\right)\frac{p\sin{\left(\phi+2\omega t\right)}}{1-rL\cos{\left(\phi+2\omega t\right)}}|t\rangle
    \end{aligned}
\end{equation}
which is equivalent to $\left|\left.\varphi_d\right\rangle\right._{pow}$, indicating that power- and signal-recycling hold equal significance in weak-value-based power improvement. Different from previous interferometric signal-recycling schemes, the only 'clockwise' path permits all cyclic photons to be refreshed prior to the last weak interaction. With the filter, the pre-filter state is $\left|\left.\varphi\prime\right\rangle\right._{sig}=M_{22}\left|\left.\varphi_0\right\rangle\right./\sqrt{\left|M_{22}\left|\left.\varphi_0\right\rangle\right.\right|^2}$, which is also equal to $\left|\left.\varphi\prime\right\rangle\right._{pow}$. Thus, the same conclusion can be obtained when considering the calculation of detected power and SNR, where $I_{sig}=I_{pow}$ and ${SNR}_{sig}={SNR}_{pow}$

\begin{figure*}[t]
\centering
\includegraphics[trim= 0 0 0 0 ,clip, scale=1.2]{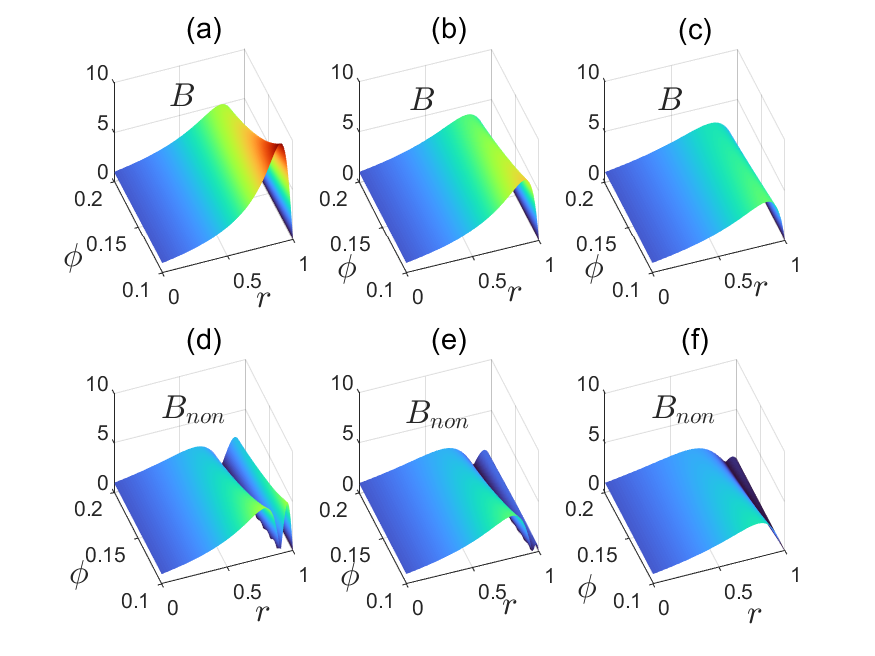}
\caption{(Color online) Comparison of polarization-based and interferometric dual recycling schemes. (a), (b) and (c) correspond to the case of $B$ varying with $r$ and $\phi$ under different values of $\gamma$: 0, 0.1 and 0.2 in 3D, respectively. (d), (e) and (f) represent that $B_{non}$ varies with $r$ and $\phi$ under different value of $\gamma$: 0, 0.1 and 0.2 in 3D. $B$ and $B_{non}$: the improvement factor of polarization-based and non-polarization dual recycling schemes. $r$: the reflection coefficient of PTM. $\phi$: post-selected angle. $\gamma$: optical loss.}
\label{Fig.3}
\end{figure*}

\subsection{Dual-recycling}
The dual-recycled WWA scheme is shown in Fig. 1(d), where the Glan prism together with two QWPs provide the input state $\left|\left.\psi_1\right\rangle\ \right.$ and the output state $\left|\left.\psi_2\right\rangle\right.$ is provided by the combination ‘HWP PBS QWPs’. In this recycling process, all possible forms of post-selections are available so that the measurement operator of per traversal can be any of $M_{11}, M_{12}, M_{21}, M_{22}$. Similarly, the filter in front of the PHWP projects the meter state into $\left|\left.\varphi_0\right\rangle\right.$, thus eliminating the walk-off effect and maintaining the large point shift associated with the WVA. This also results in a minimum optical loss $\gamma_{min}\approx4\omega^2\tau^2\phi^2$, which can be ignored in the weak value range $2\omega\tau\ll\phi\ll1$. For simple calculation, we assume the parameters of PTMs are the same and introduce the measurement matrix
\begin{equation}
U=\begin{bmatrix}M_{11}&M_{12}\\M_{21}&M_{22}\\\end{bmatrix}=\begin{bmatrix}\cos{\left(\phi+2\omega t\right)}&\sin{\left(\phi+2\omega t\right)}\\-\sin{\left(\phi+2\omega t\right)}&\cos{\left(\phi+2\omega t\right)}\\\end{bmatrix}
\end{equation}
which is formed by four measurement operators and arranged in the order corresponding to the subscripts. $\left(U^n\right)_{12}$ represents the physical process that the incident light travels through the dual recycling cavity $n$ times and finally reaches the detector. Therefore, the steady state amplitude detected by the meter is given by the sum of amplitude of all traversal numbers,
\begin{equation}
    \begin{aligned}
        &\left\langle t\right.\left|\left.\varphi_d\right\rangle\right.=p\sum_{n=0}^{n_{max}}{\left(\sqrt{1-\gamma}\right)^{n+1}\left(U^{n+1}\right)_{12}p\left\langle\alpha\right.\left|\left.\varphi_0\right\rangle\right.}\\
        &\approx p^2\left(\frac{N^2}{2\pi\tau^2}\right)^\frac{1}{4}\exp\left(-\frac{t^2}{4\tau^2}\right)\hat{L}\left(\frac{U}{I-r\sqrt{1-\gamma}U}\right)_{12}\\
        &=-\frac{p^2\left(\frac{N^2}{2\pi\tau^2}\right)^\frac{1}{4}\exp\left(-\frac{t^2}{4\tau^2}\right)\sqrt{1-\gamma}\sin{\left(\phi+2\omega t\right)}}{1+\left(1-\gamma\right)r^2-2\sqrt{1-\gamma}r\cos\left(\phi+2\omega t\right)}\\
        &\approx-\frac{p^2\left(\frac{N^2}{2\pi\tau^2}\right)^\frac{1}{4}\exp\left(-\frac{t^2}{4\tau^2}\right)\sqrt{1-\gamma}\sin{\left(\phi+2\omega t\right)}}{1+\left(1-\gamma\right)r^2-2\sqrt{1-\gamma}r\cos{\phi}},
    \end{aligned}
\end{equation}
Where the last approximation is taken with the minimum ‘filter’ loss $\gamma_{min}\approx4\omega^2\tau^2\phi^2$. In this way, the intensity of the detected signal is given by
\begin{equation}
    I_{dua}\approx\left(\frac{N^2}{2\pi\tau^2}\right)^\frac{1}{2}B^2\sin^2{\left(\phi+2\omega t\right)}\exp\left[-\frac{\left(t-\delta t\right)^2}{2\tau^2}\right]
\end{equation}
where
\begin{equation}
    B=\frac{p^2}{1+\left(1-\gamma\right)r^2-2\sqrt{1-\gamma}r\cos{\phi}}.
\end{equation}
Thus, the corresponding SNR is 
\begin{equation}
    {\rm SNR}_{dua}\approx B\frac{\sqrt N\phi}{\tau}\delta t.
    \label{e8}
\end{equation}

\section{Comparison}
Here, we define $A$ and $B$ are the improvement factors of power(or signal) and dual recycling, respectively. It is clear that the power-, signal- and dual-recycling schemes improve the SNR of standard WWA setup $A$, $A$ and $B$ times, respectively. $A^2$ and $B^2$ also corresponds to the improvement of the detected power. Therefore, as shown in Fig. 2, we plot $A$ and $B$ varying with $r$ (Fig. 2(a), (b) and (c)) or $\phi$ (Fig. 2(d), (e) and (f)) under different values of loss $\gamma=0,\ 0.1,\ 0.2$, which correspond to ideal, low and regular loss, respectively. 

As expected, the accuracy of both recycling techniques can easily exceed the corresponding standard scheme's shot noise limit, which is proportionally scaled to 1 in Figs. 2 and 3. The similar conclusions exit in \cite{32,33,34,zhong2023dualrecycled}.
However, we have to declare that it cannot beat the standard quantum limit (SQL) since the improvement is originated from the increasing of detected photons $N\phi^2 \to B^2N\phi^2$. Different cyclic schemes only stretch the standard quantum limit to varying degrees.
In addition, both $A$ and $B$ can reach the maximum value $1/\phi=10$, as shown in Fig. 2(a), (b) and (c), where the SNR itself is amplified by the large weak value factor. However, the peak of $A$ decreases faster than that of $B$, corresponding to a larger limitation to the improvement of detecting. Therefore, the dual-recycling cavity has tolerance for a wider range of $r$ and $\gamma$, which applies to more circumstances. In Fig. 2(d), (e) and (f), we set $r=0.9$, a common parameter of the PTM, and draw the curves of $A$ and $B$ varying with $\phi$ where $\phi\in\left[0.01,\ 0.2\right]$. We can see that the improvement factor of dual recycling is larger in most weak value range, thus outperforming the power or signal recycling.

In previous dual-recycled interference-based WWA system\cite{zhong2023dualrecycled}, the amplification effect of pointer is reduced by the walk-off effect, leading to a limitation of the precision gain. From the equations (25) and (26) in \cite{zhong2023dualrecycled}, without a filter, the improvement factor $B$ changes to $B_{non}$

\begin{equation}
    B_{non}=\xi\frac{p^2}{1+\left(1-\gamma\right)r^2-2\sqrt{1-\gamma}r\cos{\phi}},    
\end{equation}
where 
\begin{equation}
    \xi=\phi\frac{\cos{\phi\left[1+r^2\left(1-\gamma\right)\right]}-2r\sqrt{1-\gamma}}{\sin{\phi}[1+r^2\left(1-\gamma\right)-2r\sqrt{1-\gamma}\cos{\phi}]}.
\end{equation}
Due to the proper use of the filter, a minimum filter loss replaces original performance reduction in the polarization-based dual recycling scheme. For clear comparison, in Fig. 3, we similarly set $\phi=0.1$ and plot $B$, $B_{non}$ varying with $r$ and $\phi$ under different losses. The polarization-based scheme has obvious improvement and the gaps between $B_{non}$ and $B$ decrease as the loss increases. This is established on the assumption that both systems experience the same loss $\gamma$. Actually, replacing the BS with the PBS can effectively reduce optical loss. The probability of photons surviving the PBS ($\geq95\%$) is known to be larger than that of the BS ($\geq90\%$).  In addition, the PBS simplifies the propagation paths of cyclic photons, which reduces the crosstalk among photons. All these reasons make this polarization-based scheme advantageous in both theoretical performance and experimental application.

\section{Conclusion}

In summary, we have proposed three polarization-based cyclic weak measurement schemes based on the angular-velocity weak measurement setup. By integrating one or two PTMs into the system to establish a resonant cavity, all incident light can be detected in principle. In our analysis, these polarization-based schemes can outperform the previous interferometric schemes due to their lower theoretical loss and improved cyclic paths. 
This optimized cyclic paths effectively eliminates the walk-off effect, a significant challenge in previous signal-recycling and dual-recycling schemes. Notably, among the proposed schemes, the polarization-based dual-recycling scheme demonstrates the widest optimal region.

The application of these cyclic modes are not limited to our specific experimental setup but can be extended to various weak-value-amplification realizations. This is due to the inherent presence of post-selection in all weak value setups. In addition, postselection has been proven to improve information-cost rate, and negative quasiprobabilities enable postselected experiments to outperform optimal postselection-free experiments\cite{57}. The combination of recycling and negative quasiprobabilities represents a novel and meaningful approach. Moreover, leveraging quantum resources allows for precision enhancement beyond the standard quantum limit\cite{43,44,PhysRevLett.115.120401}, providing a predictable pathway towards further augmenting the performance of weak-value-based metrology.

\section{acknowledgements}

This work was supported by the National Natural Science Foundation of China (Grants No. 61875067).
\nocite{*}

\appendix

\section{Fisher information analysis of conventional measurement}

Considering a system prepared in $\left| R \right \rangle$, one of the basis system state. After the weak interaction, the parameter $\omega$ is encoded in the meter state as $\left | \varphi_c  \right \rangle= \left(N^2/2\pi\tau^2\right)^{1/4} \exp{(-2i\omega t)}  \exp\left(-t^2/4\tau^2\right)\left|\left.t\right\rangle\right.$. Using the quantum Fisher information (QFI) formula

\begin{equation}
    \mathcal{QF}(\omega)= 4\left[(\frac{d\left \langle \varphi \right | }{d\omega } )(\frac{d\left | \varphi  \right \rangle }{d\omega } ) - {\left |\left \langle \varphi \right| (\frac{d\left | \varphi  \right \rangle }{d\omega } ) \right |}^2 \right],
\end{equation}
we can easily get the QFI encoded in $\left | \varphi_c  \right \rangle$ as
\begin{equation}
    \mathcal{QF}_c(\omega)=16N\tau^2,
\end{equation}
which is also the maximum FI over all possible generalized measurements. 
Thus, the minimum uncertainty of $\omega$ is 
\begin{equation}
    \Delta \omega_c=\frac{1}{\sqrt{\mathcal{QF}_c(\omega)}} = \frac{1}{4\sqrt[]{N}\tau  }.
    \label{e9}
\end{equation}
The corresponding SNR is
\begin{equation}
    \rm{SNR}_c=\frac{\omega}{\Delta \omega_c}=4\omega\tau\sqrt{N}.
    \label{e10}
\end{equation}
Comparing $\Delta \omega_c$, $\rm{SNR}_c$ with $\Delta \omega$, $\rm{SNR}$ (Eqs. (\ref{e4}) and (\ref{e5})),  we can get the conclusion that standard weak-value measurement cannot offer better metrological precision of detecting the time shift of Gaussian beam, which is consistent with the analysis in \cite{PhysRevLett.114.210801}. Post-selection here acts as a concentration of the information about the parameter to be estimated into a small collected parts. 

However, by comparing Eqs. (\ref{e9}) (\ref{e10}) to Eqs. (\ref{e6}) (\ref{e7}) (\ref{e8}), it can be seen that recycled weak-value measurement can really enhance metrological precision and outperform conventional measurement. Moreover, this enhancement is consistent with the standard quantum limit (SQL) since it comes from an increase in participating photons $N \to A^2N$ (or $B^2N$). In this process, the SQL is stretched from $1/\sqrt{N}$ to $1/\sqrt{A^2N}$ (or $1/\sqrt{B^2N}$), without being exceeded.

\section{Optical mode matching and stability analysis}

Generally, the beam will be a diffracting Gaussian beam with a waist, as opposed to the parallel beam treated above.
Therefore, the waist size and positions of the incident Gaussian beam should match those of the resonance cavity itself, forming a stable self-reproduction. Here, we take the most challenging dual-recycling as an example, as illustrated in Fig. 4. 
Similar to the solutions of \cite{32,zhong2023dualrecycled}, several well-designed lenses $L_1, L_2$ and $L_3$ are employed to ensure the waists are located at PTM1 and PTM2. 

\begin{figure}[h]
\centering
\includegraphics[trim= 0 0 0 0 ,clip, scale=0.45]{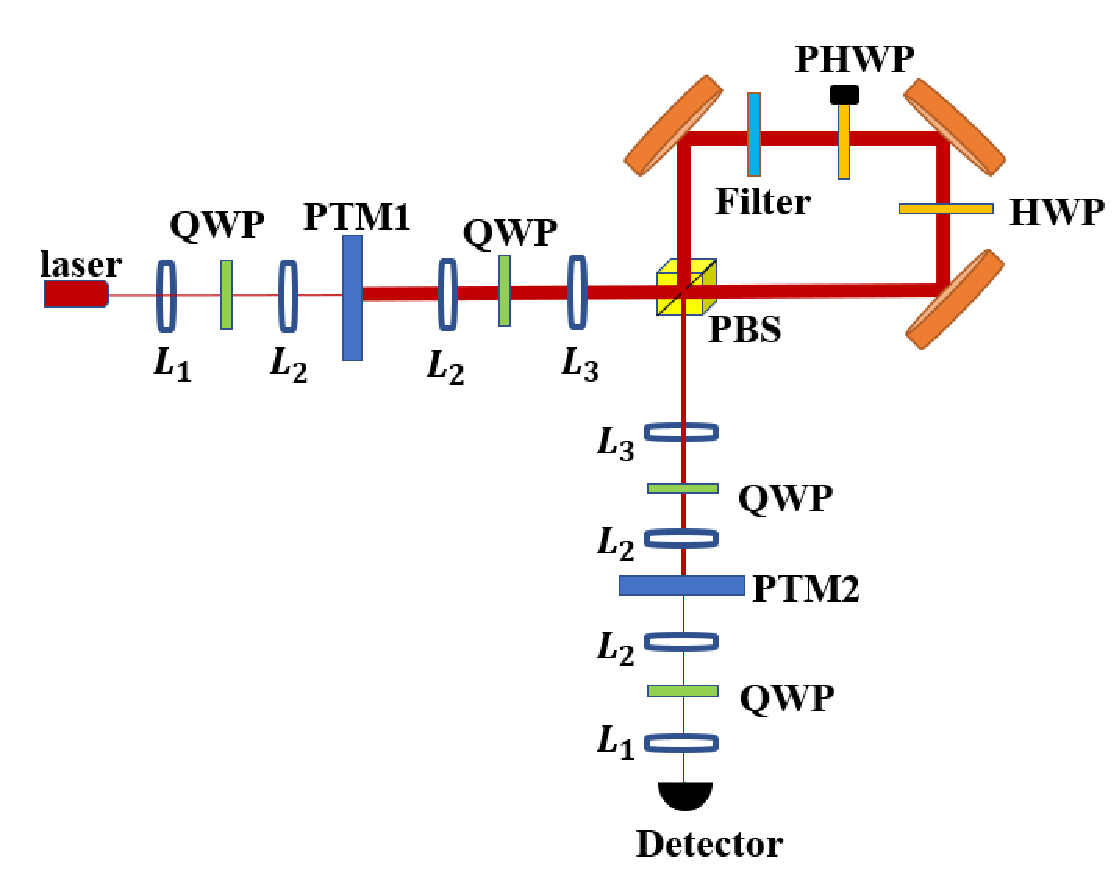}
\caption{(Color online) Optical mode matching and cavity lengths locking. The lenses $L_1$, $L_2$ and $L_3$ with matched focal lengths are set on proper positions to ensure that the place of the waist and self-reproduction of Gaussian beam are at PTM1 and PTM2. QWP: quarter-wave plate. HWP: half-wave plate. PBS: polarization-beam-splitter. PHWP: piezo-driven half-wave plate. PTM: partially transmitting mirror.}
\label{Fig.4}
\end{figure}

The phase locking of the cavity is also essential since the ambient noise can introduce length offsets in the cavity. In Ref. \cite{32}, an error signal extracted from the output light is used to provide feedback and stabilize the power-recycling cavity using the Pound-Drever-Hall (PDH) method, which is also applicable to the signal-recycling cavity. However, when implementing the dual-recycling cavity with two different lengths, relying solely on an error signal does not provide sufficient information regarding the offsets of the PTMs. A possible approach is to adjust the post-selected angle and parameters of the PTMs such that the light reflected towards the laser is largely independent of PTM2, allowing the first PDH system to initially lock the length of PTM1 before using the second PDH system to stabilize the length of PTM2. Alternatively, a custom-designed enclosure with fixed positions for the instruments can simplify the locking operations.

\end{document}